\documentclass[%
 reprint,
 amsmath,amssymb,
 aps,
]{revtex4-1}

\usepackage{dcolumn}
\usepackage{bm}
\usepackage{color}

\usepackage{amssymb,amsmath}
\usepackage{bm}

\usepackage[normalem]{ulem}

\usepackage{bm}
\usepackage{multirow}
\usepackage{mathrsfs}
\usepackage{graphicx}
\usepackage{epstopdf}

\usepackage{soul}

\usepackage[caption=false]{subfig}
\usepackage{epstopdf, epsfig}

\usepackage{float}

\newcommand{\feq}{f^{\rm eq}}
\newcommand{\fmirr}{f^{\rm mirr}}

\newcommand{\edotshift}[2]{\left\langle {#1} \mid {#2} \right\rangle^\prime}

\begin{document}

\preprint{APS/123-QED}

\title{Fluid-Structure Interaction with the Entropic Lattice Boltzmann Method}

\author{B. Dorschner}
  \email{benedikt.dorschner@lav.mavt.ethz.ch}
\author{S.S. Chikatamarla}%
 \email{chikatamarla@lav.mavt.ethz.ch}
 \author{I.V. Karlin}%
 \email{Corresponding author; karlin@lav.mavt.ethz.ch}
\affiliation{Aerothermochemistry and Combustion Systems Lab, Department of
    Mechanical and Process Engineering, ETH Zurich, CH-8092 Zurich, Switzerland}

\date{\today}

\begin{abstract}

We propose a novel fluid-structure interaction (FSI) scheme using the entropic 
multi-relaxation time lattice Boltzmann (KBC) model for the fluid domain in
combination with a nonlinear finite element solver for the structural part. 
We show validity of the proposed scheme for various challenging set-ups by
comparison to literature data.
Beyond validation, we extend the KBC model to multiphase
flows and couple it with FEM solver.
Robustness and viability of the entropic multi-relaxation time model for complex
FSI applications is shown by simulations of droplet impact on elastic superhydrophobic surfaces.

\end{abstract}

\maketitle


\section{Introduction}

Fluid-structure interaction (FSI) is of significant interest in science and
engineering applications, where examples include aeroelasticity such
as flutter and buffeting 
\cite{Vu2016,Beskhyroun2011,Farhat2006,Barlas2010},
or bio-fluidmechanics in order to enhance our understanding of 
cell aggregation, blood-heart interaction as well as 
the propulsion mechanisms in flying and swimming
\cite{Kern2006,Shyy2010,Ristroph2010,Novati2017,Bottom2016,Whitehead2015}.
Insight in these phenomena through experimental and numerical studies has shown
tremendous success for example, the development of cancer diagnostic
devices the size of a chip \cite{Sackmann2014,Krueger2014}, optimizing
wind turbines \cite{Barlas2010} or artificial heart valves
\cite{Borazjani2010} but also may be used to draw
inspirations for the design of novel propulsion system in robotic devices. 
%
Yet, such problems remain a challenge to existing methods due to strong
nonlinearity and their multidisciplinary nature \cite{Dowell2001}.
Due to inherent limitations of experiments in terms of accuracy, accessibility
and cost much research effort has been devoted to the development of numerical
methods for the simulation of FSI applications.

In general, there are two main avenues to FSI simulations, namely 
monolithic and partitioned approaches.  The monolithic approach describes the
fluid and the structural part with the same discretization scheme in one system
of equations, which is solved simultaneously with a single solver.  This
technique implies consistent fluid-structure interface conditions.  On the other
hand, in most practical applications the partitioned approach is employed, which
uses separate solvers for the fluid and structural part, respectively.
The advantage of this approach is its modularity, which allows the use of
independently optimized solution strategies in the solid and fluid domain,
respectively.  Thus partitioned approaches are most common and therefore focus
of this paper.  On the other hand, consistent solid-fluid interface conditions 
are not satisfied implicitly and thus pose the main challenge of partitioned
fluid-structure approaches. 
For the simulation of the fluid flow involving complex moving geometries, a
broad categorization into body-conforming and non-conforming methods can be
identified. Most conventional FSI schemes are based on body-fitted grids, where
the interface conditions are treated as boundary conditions and the
computational mesh conforms to the moving and deforming solid-fluid interface.
Examples of commonly used body-fitted approaches include
arbitrary-Lagrangian-Eulerian formulations \cite{Hirt1974,Ahn2006} and
space-time finite element methods \cite{TEZDUYAR1992,Tezduyar2006}. While the
interface conditions are easily imposed, the generation of the moving meshes for
complex geometries undergoing large deformations is computationally expensive
and requires sophisticated procedures to avoid severe mesh distortion to
preserve accuracy
\cite{thompson1998handbook,Hermansson2003,Tezduyar2006,Nakata2012}. Viable
alternatives are found in non-conforming methods, which we will focus on in this
paper.

In particular, we employ the lattice Boltzmann method (LBM), which has matured
into an attractive alternative to conventional methods based on a direct
discretization of the macroscopic Navier-Stokes equations. The LBM derives from
kinetic theory and evolves discretized particle distribution functions
(populations) $f_i(\bm x, t)$, which are associated with a set of discrete
velocities $\bm c_i=1, \cdots, Q$ and designed to recover the macroscopic
Navier-Stokes equations in the hydrodynamical limit. By organizing $\bm c_i$
into a regular lattice, the LBM reduces to a propagation step, advecting the
populations along $\bm c_i$, and a collision operator, which over-relaxes the
populations. In this paper, we chose the entropic multi-relaxation time
collision operator \cite{Karlin2014} for its accuracy and robustness as shown in
various challenging set-ups involving a combination of turbulence and complex
moving geometries \cite{Dorschner2016}.  For modeling complex, moving geometries
we employ Grad's approximation as proposed in \cite{Dorschner2015}, which has
shown to be reliable for both one- and two-way coupled simulations of moving and
deforming objects \cite{dorschner2017entropic}.  \\ The structural domain on the
other hand is described by a geometrically nonlinear total Lagrangian
formulation in the framework of the finite element method (FEM).  \\ In this
paper, we aim to asses the predictive capabilities of the recently developed
entropy-based lattice Boltzmann models in combination with Grad's approximation
for fluid-structure interaction problems involving large deformations.  Beside
thorough validation by comparison to standard benchmarks, the robustness of the
scheme allows us to explore its capabilities in multi-physics applications,
where we present a novel multiphase formulation of the entropic multi-relaxation
time model and its coupling to the structural solver.

The paper is structured as follows: We begin by briefly introducing the
numerical methodology to solve the governing equations for both the fluid and
the solid domain, followed by a discussion of their coupling through appropriate
boundary conditions.
In section \ref{sec:results}, we report the numerical results obtained by the
proposed scheme. We start by a thorough validation of the model in section
\ref{sec:validation} and subsequently present the extensions of the FSI
scheme for multiphase flows in section \ref{sec:fsi_mpf}.

\section{Numerical approaches}
In the following we briefly describe the numerical methodology for the fluid,
the elastic solid as well as the coupling methodology.  The subscript $f$ and
$s$ are used to indicate the fluid and structural quantities, respectively.  The
time-dependent fluid and solid domain with their common interface are denoted by
$\Omega_{f}^t$, $\Omega_{s}^t$ and $\Gamma_I^t=\partial \Omega_f^t \cap \partial
\Omega_s^t$, respectively.  The corresponding reference or initial domains and
the interface are referred to as $\Omega_{s}$, $\Omega_{s}^t$ and
$\Gamma_I=\partial \Omega_f \cap \partial \Omega_s$, respectively.  Further,
Neumann and Dirichlet boundary conditions are identified as $\Gamma_N$ and
$\Gamma_D$, respectively.

\subsection{Entropic multi-relaxation time lattice Boltzmann method}
We solve the weakly compressible Navier-Stokes equations using the entropic
multi-relaxation time lattice Boltzmann (KBC) model.  For brevity, we only summarize
the main steps and refer to the works of \cite{Karlin2014, Bosch2015,
Dorschner2016} for a thorough discussion of the model.  The evolution of the
population  $f_i(\bm x,t)$ is given by the discrete kinetic equation 
\begin{equation}
\label{eq:f_equations}
	f_i(\bm{x+c_i}, t+1)=f_i^{\prime} = (1-\beta)f_i(\bm{x},t) + \beta f_i^{\text{mirr}}(\bm{x},t),
\end{equation}
where the advection step is indicated by the left-hand side and the
post-collision state $f^\prime_i$ is represented on the right-hand side.  The
populations are expressed in its natural moment basis as the sum of the
kinematic part $k_i$, the shear part $s_i$ and the remaining higher-order
moments $h_i$ as
\begin{equation}
	f_i = k_i + s_i + h_i.
\label{eq:popSplit}
\end{equation}
The mirror state may thus be defined as 
\begin{equation}
\label{eq:f_mirr}
	f_i^{\text{mirr}} = k_i + \left( 2 s_i^{\rm eq} -s_i \right) + \left( \left(1 - \gamma\right)  h_i + \gamma h_i^{eq}\right),
\end{equation}
where $s_i^{eq}$ and $h_i^{eq}$ are $s_i$ and $h_i$ evaluated at equilibrium
conditions.  The equilibrium $f^{\rm eq}$ is defined as the minimum of the
entropy function
\begin{equation}
	H(f) = \sum_{i=1}^Q f_i \ln \left( \frac{f_i}{W_i} \right) , \\
	\label{eq:feq_min}
\end{equation}
subject to the local conservation laws for mass and momentum
\begin{equation}
	\sum_{i=1}^Q \lbrace 1, \bm{c_i} \rbrace f_i = \lbrace \rho, \rho \bm{v}_f \rbrace,
	\label{eq:conservationLaws}
\end{equation}
where the weights $W_i$ are lattice-specific constants. 

Finally the relaxation rate $\gamma$ of the higher-order moments is found by
minimizing the discrete entropy function (see Eq.~\ref{eq:feq_min}) in the
post-collision state $f_i^\prime$.  Thus, at every time step and every grid
point the estimate for $\gamma$ is computed by the following analytical
expression 
\begin{equation}
	\gamma = \frac{1}{\beta} - \left( 2 - \frac{1}{\beta}\right) \frac{\left< \Delta s | \Delta h\right>}{\left< \Delta h | \Delta h\right>  },
	\label{eq:gamma_min_approx}
\end{equation}
where $\Delta s_i = s_i - s_i^{\text{eq}}$, $\Delta h_i=h_i-h_i^{\text{eq}}$ and 
$\left< X | Y \right> = \sum_i (X_i Y_i / f_i^{\text{eq}})$.

The entropic multi-relaxation time lattice Boltzmann method recovers the
Navier-Stoke equations in the hydrodynamic limit and relates the parameter
$\beta$ to the kinematic viscosity as
\begin{equation}
	\nu=c_s^2 \left( \frac{1}{2\beta}- \frac{1}{2}\right),
	\label{eq:visc}
\end{equation}
where $c_s=1/\sqrt{3}$ is the lattice speed of sound.

\subsection{Structural modeling}
In the Lagrangian frame, the structural part is governed by the momentum equation  as
\begin{align}
\begin{split}
\rho_s \frac{\partial \bm v_s}{ \partial t }  - \nabla \cdot \bm P_s &= \rho_s
\bm b_s, \quad \text{in } \Omega_{s} \\
\bm v_s  &= \overline{\bm v}_s, \quad \text{on } \Gamma_{D} \\
\bm P_s \bm n_s  &= \overline{\bm t}_s, \quad  \text{on } \Gamma_{N}
\end{split}
\label{eq:strongForm}
\end{align}
where $\bm v_s$, $\rho_s$ and $\bm b_s$ are the solid velocity, density and body
force per unit mass. The outer normal vector of $\Gamma_I$ or $\Gamma_{s,N}$ is
denoted by $\bm n_s$. The prescribed velocities and tractions on the Dirichlet
and Neumann boundary are indicated by $\overline{\bm v}_s$ and $\overline{\bm
t}_s$, respectively.  The first Piola-Kirhoff stress is denoted by $\bm P_s$ and
related to the second Piola-Kirhoff stress $\bm S_s$ by 
\begin{equation}
\bm P_s=\bm F \bm S_s,
\end{equation}
where $\bm F$ denotes the deformation gradient 
\begin{equation}
\bm F=\bm I + \nabla \bm u_s
\end{equation}
and $\bm u_s$ is the displacement field of the solid. The second Piola-Kirhoff
stress on the other hand can be mapped to the Cauchy stress tensor $\bm
\sigma_s$ by
\begin{equation}
\bm S_s=J \bm F^{-1} \bm \sigma_s \bm F^{-T},
\end{equation}
where $J=\det(\bm F)$.
In this paper, we consider the hyperelastic Saint Venant-Kirchoff constitutive
equation, which extents linear elastic models to the geometrically nonlinear
regime and defines the second Piola-Kirhoff stress as
\begin{equation}
\bm S_s=\lambda tr(\bm E) \bm I + 2 \mu_s \bm E,
\end{equation}
where 
\begin{equation}
\bm E= \frac{1}{2} ( \bm F^T \bm F - \bm I ) =  \frac{1}{2}( \nabla \bm u_s + \nabla \bm u_s^T + \nabla \bm u_s^T \nabla \bm u_s  )
\end{equation}
is the Green-Lagrangian strain tensor.  The first and second Lam\'e coefficients
are indicated by $\lambda_s$ and $\mu_s$, respectively and are related to
Young's modulus $E_s$ and Poisson's ratio $\nu_s$ as
\begin{equation}
\nu_s = \frac{\lambda_s}{2 (\lambda_s +  \mu_s)},  \quad E_s=\frac{ \mu_s(3\lambda_s+2\mu_s)}{\lambda_s+\mu_s}.
\end{equation}
In the present work, we employ a two-field formulation and solve for the
displacement field separately using the kinematic compatibility condition
\begin{align}
\begin{split}
\frac{\partial \bm u_s}{\partial t} - \bm v_s &=0 \quad \text{in } \Omega_s \\
\bm u_s  &= \overline{\bm u}_s, \quad \text{on } \Gamma_{D},
\end{split}
\end{align}
where $\overline{\bm u}_s$ denotes the prescribed displacement on the Dirichlet
boundary.

The structural equations are solved using the finite element method (FEM), which
is implemented in the open-source library deal.ii \cite{Bangerth2007}.  We
follow standard FEM procedures, see, e.g., the textbooks \cite{Kim2015,Bathe} or
in the context of monolithic FSI with deal.ii
\cite{Wick2011a,Wick2011,Richter2012}.
Using the conventional notation for Lebesgue and Sobolev spaces, we define the
following functional spaces for trial and weighting functions:
\begin{align}
    &\mathcal{L} :=\lbrace \bm w_s \in L^2(\Omega_s) \rbrace, \\
    &\mathcal{V}_0 := \lbrace \bm w \in H^1(\Omega_s): \bm w=0 \text{ on } \Gamma_{s,D} \subset
    \Omega_s \rbrace  \\
    &\mathcal{V}_D := \lbrace \bm w \in H^1(\Omega_s): \bm w= {\bm {w}}_{s,D} \text{ on } \Gamma_{s,D}
    \subset \Omega_s \rbrace 
\end{align}
where $L^2$, $H^1$ denote the Lebesgue space of square integrable functions and
the first Sobolev space, respectively.  Furthermore, the short-hand notations
$(\cdot, \cdot )$ and $\langle \cdot, \cdot \rangle$ indicate the scalar product
on the $L^2$-space and its boundary, respectively.  Thus, following standard
procedures, we obtain the following variational formulations for
$ \lbrace \bm v_s, \bm u_s \rbrace \in \lbrace \mathcal{L} \times \mathcal{V}_D \rbrace$
\begin{align} \label{eq:weakForm_solid0}
   (\rho_s \partial_t \bm v_s, \bm \psi_{s,v})_{\Omega_s}  + 
  ( \bm P_s, \nabla \bm \psi_{s,v})_{\Omega_s} - 
   (\rho_s \bm b, \bm \psi_{s,v})_{\Omega_s} - &
\nonumber
\\
  \langle \bm t , \bm \psi_{s,v} \rangle_{\Gamma_I \cup \Gamma_{s,N}}
    = 0 \quad 
    \forall \bm \psi_{s,v} \in \mathcal{V}_0, &\\
%
  (\partial_t \bm u_s, \bm \psi_{s,u})_{\Omega_s}  -
   ( \bm v_s, \bm \psi_{s,u} )_{\Omega_s} =0 
    \quad \forall \bm \psi_{s,u} \in \mathcal{L}, &
\end{align}
where $\psi_{s,u}$, $\psi_{s,v}$, $\bm u_s$ and $\bm v_s$ are the trial and
test functions of the solid displacement and velocity, respectively.
Note that the traction $\bm t$ may also be specified in terms of the Cauchy
stress tensor 
$ \bm \sigma_s$ as 
\begin{equation}
    \bm t = J_s \bm \sigma_s \bm F_s^{-T} \bm n_s
\end{equation}
For simplicity, we use the one step-$\theta$ scheme 
for the integration in time, which, for a
generic quantity $g$ with $\partial_t g(t) = f(t, g(t))$, reads
\begin{equation}
\partial_t g \approx \frac{g^{n+1}- g^{n}}{\Delta t} = \theta f^{n+1} + (1- \theta ) f^n.
\label{eq:timeInt}
\end{equation}
This allows us to choose implicit/explicit Euler or centered/shifted
Crank-Nicolson time integration depending on the choice of $\theta$ 
but can also easily be extended to the fractional-step-$\theta$ scheme.
Note that the following can be extended in a straightforward manner to other
standard time integration schemes such as the Newmark algorithm or alike.

Using the temporal discretization of Eq.~\eqref{eq:timeInt}, the variational
formulation of Eq.~\eqref{eq:weakForm_solid0} may be discretized in time as 
\begin{align} \label{eq:timeDiscrete_solid_velocity}
&   \rho_s \Delta t^{-1} \left( \bm v_s^{n+1}, \bm \psi_{s,v}\right)_{\Omega_s}  + 
  \theta \left( \bm P_s^{n+1}, \nabla \bm \psi_{s,v}\right)_{\Omega_s} = 
\\ \nonumber
&   \rho_s \Delta t^{-1} \left( \bm v_s^{n}, \bm \psi_{s,v}\right)_{\Omega_s}  + 
    \theta 
    \left[  
       \langle \bm t^{n+1}, \psi_{s,v}\rangle_{\Gamma_I \cup \Gamma_{s,N}},   
       \left( \rho_s \bm b^{n+1}, \psi_{s,v}\right)_{\Omega_s}
   \right] 
   +
\\ \nonumber
&
    \left( 1-\theta \right)
    \left[
        \langle \bm t^{n}, \bm \psi_{s,v}\rangle_{\Gamma_I \cup \Gamma_{s,N}} - 
        \left(\rho_s \bm b^{n}, \bm \psi_{s,v}\right)_{\Omega_s} - 
        \left(\bm P_s^n, \bm \psi_{s,v}\right)_{\Omega_s}
    \right]
\\ \nonumber
&    \forall \bm \psi_{s,v} \in \mathcal{V}_0, 
\\
%
&  \Delta t^{-1} \left( \bm u_s^{n+1}, \bm \psi_{s,u}\right)_{\Omega_s}  -
  \theta \left( \bm v^{n+1} , \bm \psi_{s,u} \right)  =
\nonumber \\ &
   \Delta t^{-1} \left( \bm u_s^{n}, \bm \psi_{s,u}\right)_{\Omega_s}  -
   \left( 1- \theta \right)
   ( \bm v_s^n, \bm \psi_{s,u} )_{\Omega_s} =0 
    \quad \forall \bm \psi_{s,u} \in \mathcal{L}, 
\label{eq:timeDiscrete_solid_displacement}
\end{align}

With slight rearrangement it should be obvious that Eq.~\eqref{eq:timeDiscrete_solid_velocity} and
Eq.~\eqref{eq:timeDiscrete_solid_displacement} can conveniently be expressed in
matrix form as
\begin{equation}
\bm A( \bm U^{n+1}, \bm \Psi)= \bm F(\bm \Psi) 
\label{eq:matrixForm},
\end{equation}
where $\bm U^{n+1}=\lbrace{ \bm v_s^{n+1}, \bm u_s^{n+1} \rbrace}$ and 
$\bm \Psi=\lbrace{ \bm {\psi}_{v,s}, \bm{\psi}_{u,s}\rbrace}$.

Based on the time-discrete equations shown above, we employ a finite element
Galerkin discretization in space.  
We discretize the undeformed or reference domain $\Omega_s$ in a
shape-regular mesh $\mathcal{M}_h$, which is composed of hexahedral elemets
$\mathcal{E}$. 
The finite element spaces are given by
\begin{align}
\mathcal{L}_h &:=  \lbrace \bm w_h \in C(\Omega_h), 
                    \bm w_h \lvert_{\mathcal{E}} \in Q_p(\mathcal{E}) 
                    \quad \forall \mathcal{E} \in \mathcal{M}_h, \subseteq L^2, \\
\mathcal{V}_{0,h} &:=  \lbrace \bm w_h \in C(\Omega_h), 
                    \bm w_h \lvert_{\mathcal{E}} \in Q_p(\mathcal{E}) 
                    \quad \forall \mathcal{E} \in \mathcal{M}_h, \\
                &    \bm w_h =0 \text{ on } \Gamma_{s,D,h} \rbrace \subseteq H^1, \\
\mathcal{V}_{D,h} &:=  \lbrace \bm w_h \in C(\Omega_h), 
                    \bm w_h \lvert_{\mathcal{E}} \in Q_p(\mathcal{E}) 
                    \quad \forall \mathcal{E} \in \mathcal{M}_h, \\
                &    \bm w_h = \bm w_{s,D,h} \text{ on } \Gamma_{s,D,h} \rbrace \subseteq H^1,
\end{align}
where $Q_p(\mathcal{E})$ is the space of tensor product polynomials of to degree $p$. 
In the following, we restrict ourselves to the $Q_2$ element for simplicity, but
it can straightforwardly be extended to higher order.  Further, a bilinear
transformation is used to map the physical elements to the unit element. 

Finally, the fully time- and space-discrete nonlinear system in
matrix notation reads as 
\begin{equation}
\bm A( \bm U_h^{n+1}, \bm \Psi_h)= \bm F (\bm \Psi_h)
\label{eq:matrixForm2},
\end{equation}
for $\bm U_h^{n+1} = \lbrace \bm v_{s,h}^{n+1} \bm u_{s,h}^{n+1} \rbrace \in \lbrace \mathcal{L}_h \times
\mathcal{V}_{D,h}\rbrace$ and $\bm \Psi_h =\lbrace \bm \psi_{s,v,h},
\psi_{s,u,h}  \rbrace \in \lbrace \mathcal{V}_{h,0} \times \mathcal{L}_h \rbrace$.

The non-linear equations arising from the integration procedures and the
Saint-Venant Kirchoff constitutive relation are solved using a Newton-Raphson
method in combination with a simple line search algorithm.  This yields the
incremental updating rule for the $k$-th iteration as 
\begin{align}
\bm A^\prime ( \bm U_h^{n,k})(\delta \bm U_{n,k}, \bm \Psi_h)& = - \bm A(\bm
        U_h^{n,k}),\bm \Psi_h) + \bm F(\bm \Psi)
\\
\bm U_h^{n,k+1} &= \bm U_h^{n,k} + \lambda \delta \bm U_h^{n,k},
\end{align}
where $\lambda \in (0, 1]$ is the line search relaxation parameter.  For all
cases in this paper $\lambda =0.7$ has proven to be a good choice.  The
G\^ateaux derivatives $\bm A^\prime ( \bm U)(\delta \bm U_{n,k}, \bm \Psi_h)$
are analytically computed.  In particular, the non-linearity arises due to the
Saint-Venant Kirchoff relation, which only depends on the displacement.  Thus,
for direction $\delta \bm U$ the corresponding derivatives with respect to $\bm
U$ may be identified as
\begin{equation}
\partial_{ \bm U }\bm E = \frac{1}{2} \left( \nabla \delta \bm U \bm F +  \bm F^T \nabla
    \delta \bm U \right),
\end{equation}
which yields 
\begin{align}
\partial_{ \bm U} \bm S =& \frac{1}{2} \lambda tr \left( \nabla \delta \bm U^T \bm F +
    \bm F^T \nabla \delta \bm U \right) \bm I + 
\nonumber \\ 
&    \mu_s \left( \nabla \delta \bm U \bm F + \bm F^T \nabla \delta \bm U \right),
\end{align}
and upon substitution
\begin{align}
\bm A^\prime ( \bm U)(\delta \bm U, \bm \Psi_h) =&
\left(
    \nabla \delta \bm U \bm S +  
\right.
\nonumber \\  
    &\bm F \left(
        \mu_s (\nabla \delta \bm U^T \bm F +  \bm F^T \nabla \delta \bm U ) + 
        \right.
\nonumber \\
&\left.
        \left.
        \lambda tr ( \bm F^T  \nabla \delta \bm U) \bm I 
       \right),
     \nabla \bm  \Psi_h
\right).
\end{align}

\subsection{Fluid-structure coupling}

A consistent coupling of the fluid and structural domain 
enforces the following interface conditions
\begin{align}
\bm v_f &= \bm v_s \quad \text{ on } \Gamma_I^t,
\\
\bm P_s \bm n_s + J \bm \sigma_f \bm F^{-T} \bm n_s &=0 \quad  \text{ on } \Gamma_I,
\end{align}
where the 
$\bm \sigma_f  = -p_f \bm I +  \rho_f \nu_f ( \nabla \bm v_f  + \nabla \bm v_f^T)$ 
is the fluid stress tensor.

Within the context of partitioned approaches, one can distinguish between
weakly(loose)- and strongly-coupled FSI schemes.  While weakly-coupled methods
do not enforce the fluid-solid interface constraints, strongly-coupled methods
typically utilize subiterative schemes to converge to the solution of the
monolithic system.  Weakly-coupled methods are computationally less expensive
but have shown to generate artificial energy at the interface due to the
staggered nature of the evolution of the fluid and structural part
\cite{Piperno2001}. This so-called added-mass effect can cause fatal
instabilities for small solid-fluid density ratios and thus may limit their
range of applicability
\cite{Causin2005,Borazjani2008a,LeTallec2001,Forster2007,Li2017}.  

However, the added-mass effect is proportional to the time step size for
compressible flows, and convergences to a non-zero value  only in the fully
incompressible regime.  Thus, for the weakly compressible LBM at the
incompressible limit and the corresponding small time step size this effect has
only a limited influence \cite{Kollmannsberger2009}.

Hence, for simplicity, we chose a weakly-coupled partitioned approach using the
conventional serial staggered (CSS) approach.  The fluid is solved by the LBM
and the solid by an appropriate finite element discretization, which accounts
for geometric nonlinearity.  The coupling between both domains is achieved
through appropriate boundary conditions.  On one hand, the Grad boundary
condition accounts for the coupling from the solid to the fluid. On the other
hand, the fluid is coupled to the solid by the traction force as computed
through a pressure tensor extrapolation scheme similar to
\cite{Kollmannsberger2009}. 

Thus, in the CSS algorithm, we first perform a fluid step (including boundary
conditions) and compute the force (traction) on the solid. Subsequently, the
structural solver computes its deformation, where the traction is imposed as a
boundary condition.  Finally, we transfer velocity and displacement of the solid
to the fluid solver, update the solid geometry in the fluid solver and
incorporate the boundary velocity in the fluid boundary conditions.

In the following, we briefly summarize the implementation of the corresponding
boundary conditions needed to perform full coupling.

\subsubsection{Fluid boundary conditions}

In the FSI simulations, the fluid boundary condition imposes the no-slip
condition and accounts for the momentum exerted from the solid to the fluid.  In
the realm of LBM many variants exist in literature ranging from simple
bounce-back schemes to implicitly corrected immersed boundary methods. However,
only a few have shown to be stable, accurate and universally applicable.  A
viable alternative was proposed in \cite{Dorschner2015} using an analog of
Grad's approximation for the missing populations.  This boundary condition was
shown to be second-order accurate and consistent with the entropy-based LBM.
Accuracy and robustness has been demonstrated for various challenging set-ups
involving complex, moving and deforming geometries in one- and two-way coupled
simulations in both laminar and turbulent flows
\cite{Dorschner2016,dorschner2017transitional,dorschner2017entropic}.

The Grad approximation is a parametrization of the population in terms of its
moments and in the athermal case it has shown to be sufficient to include the
pressure tensor $\bm{\Pi}$ besides the conserved quantities.  An explicit
expression is given by
\begin{align} \label{eq:grad}
	f_i^{\ast}(\rho, \bm{v}_f, \bm{\Pi}) = W_i \lbrack & \rho + \frac{\rho }{c_s^2} \bm{c_i}\cdot \bm{v}_f + \\ \nonumber
                                                    &\frac{1}{2 c_s^4} \left( \bm{\Pi} - \rho c_s^2 \bm{I} \right) :
												\left(  \bm{c_{i}} \otimes \bm{c_{i}} - c_s^2 \bm{I} \right) \rbrack ,
\end{align}
where the pressure tensor $\bm{\Pi}$ is approximated by 
\begin{equation}
\bm{\Pi}=\bm{\Pi}^{\text{eq}}+\bm{\Pi}^{\text{neq}},
\end{equation}
with 
\begin{align}
\bm{\Pi}^{\text{eq}}  &= \rho c_s^2 \bm{I}+ \rho \bm{v}_f \otimes \bm{v}_f, \\
\bm{\Pi}^{\text{neq}} &= -\frac{\rho c_s^2}{2 \beta}   \left(   \nabla \bm{v}_f +   \nabla \bm{v}_f^\dag  \right).
\end{align}

Thus, by appropriately specifying density $\rho$, velocity $\bm v_f $ and the
pressure tensor  $\bm \Pi$ we account for the momentum exerted from the fluid to
the solid as well the mass swept by the object.  For this purpose, the concept
of target values was introduced, where the details are discussed at length in
our previous contributions \cite{Dorschner2015, dorschner2017entropic} and we
will only highlight important FSI specifics here.  While the target density is
given by the implied bounce-back density with an additional contribution in
order to account for the object motion, the velocity gradients are evaluated
using a finite difference scheme.  The velocity at the locations of the Eulerian
fluid mesh on the other hand involves interpolation of the velocity from the
intersection point $\bm x_{w,i}$ with the boundary along the lattice vector $\bm
c_i$.

To that end, in the case of FSI, we use the FEM mesh to construct a surface
mesh, which is passed to the fluid solver.  In particular, we partition the
quadrilateral surface elements of the FEM solver into triangular elements for an
efficient detection of the intersection location and update the vertex locations
using the displacements as computed by the FEM solver.  Furthermore, the
corresponding velocity values are transferred to the fluid solver and used to
interpolate the velocity values at the intersection locations $\bm x_{w,i}$.
This completes the fluid boundary condition.

\subsubsection{Solid boundary conditions}
For the coupling of the fluid to the solid, we impose a traction boundary
condition as 
\begin{equation}
\bm t = J_s \bm \sigma_f \bm F_s^{-T} \bm n_s.
\label{eq:solidBC}
\end{equation}
Thus, we need to evaluate $ \bm \sigma_f$ at the quadrature points of the FEM
mesh. 
Fortunately, in LBM the fluid stress tensor can conveniently computed as 
\begin{equation}
\bm \sigma_f = -p \bm I  - (1- \beta ) \bm \Pi^{(1)}, 
\end{equation}
where $p= \rho c_s^2$ in the athermal case and 
$\bm \Pi^{(1)} = \sum_i f_i^{(1)} \bm c_i \otimes \bm c_i$, which is
evaluated using $f^{(1)}_i \approx f_i - f_i^{eq}$. 
As $\bm \sigma_f$ needs to be evaluated at all quadrature point on the solid
surface mesh, we use an extrapolation scheme, similar to
\cite{Kollmannsberger2009}.

\section{Numerical results}\label{sec:results}

\subsection{Validation} \label{sec:validation}
\subsubsection{Turek Benchmark}

\begin{figure}
        \centering
        \vspace{0.2cm}
                \includegraphics[width=0.5\textwidth]{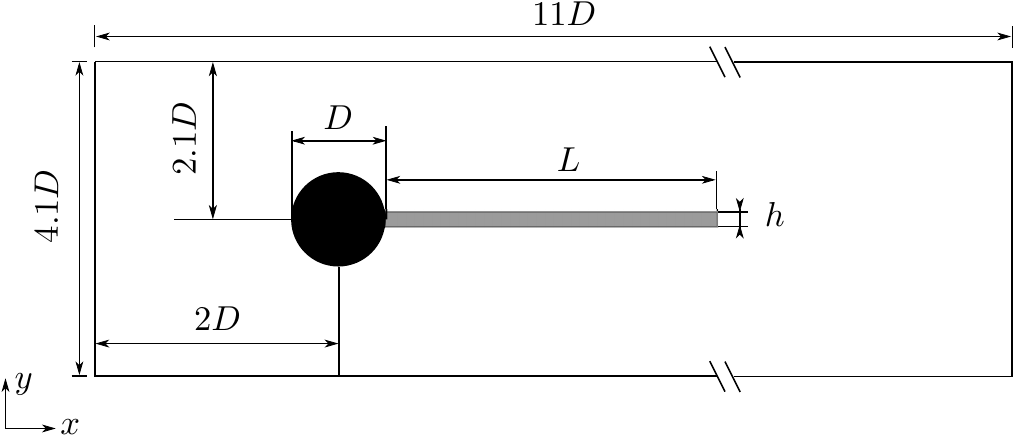}
        \caption{Schematic - Turek Benchmark.}
	\label{fig:csm3_schematic}
\end{figure}

\begin{figure}
	\centering					
	\includegraphics[width=0.45\textwidth]{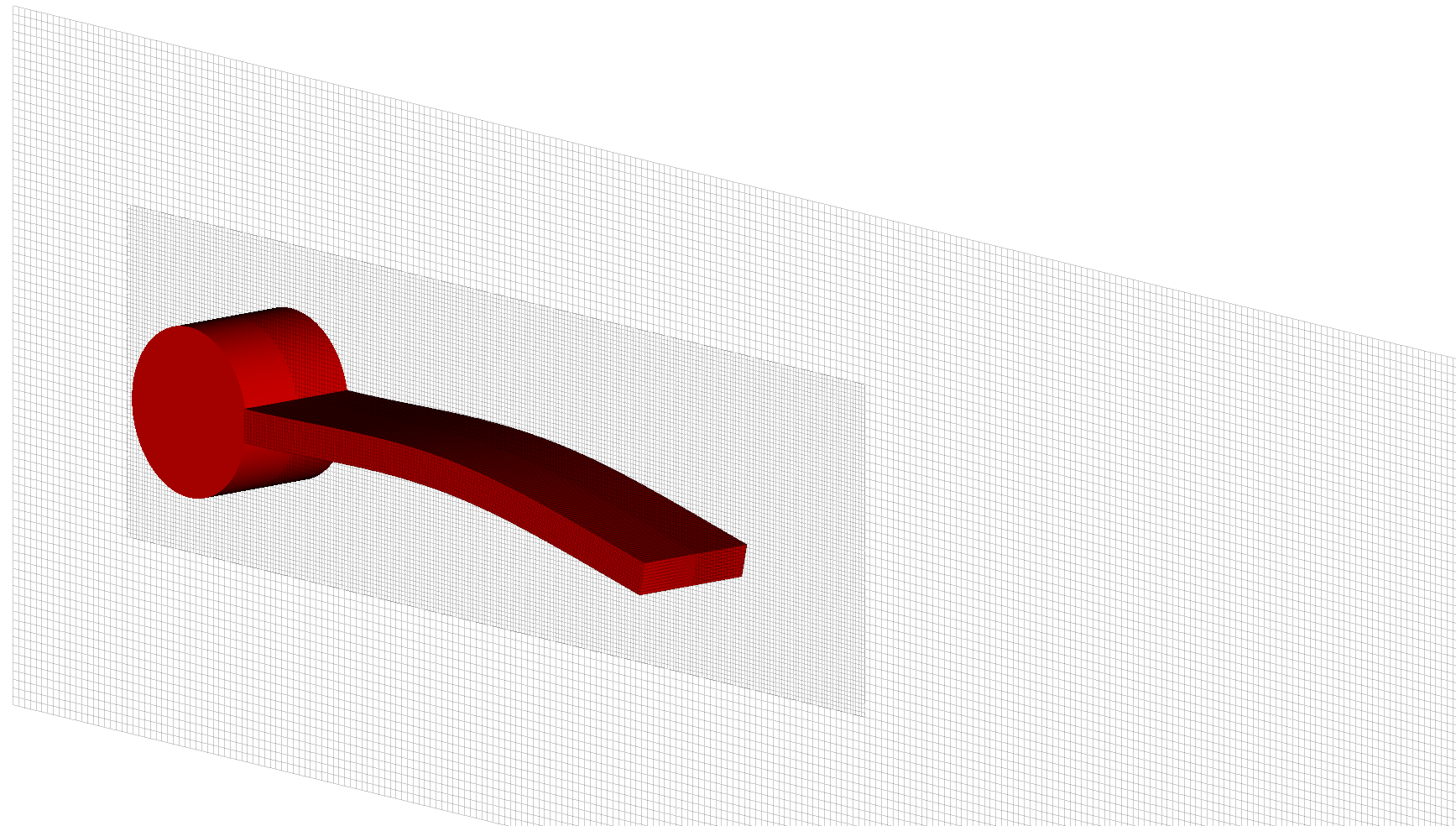}
	\caption{Snapshot of the computational domain, zoomed in on the cylinder-flag assembly.}
	\label{fig:TurekMesh}
\end{figure}
\begin{figure*}
        \centering
                \includegraphics[width=0.45\textwidth]{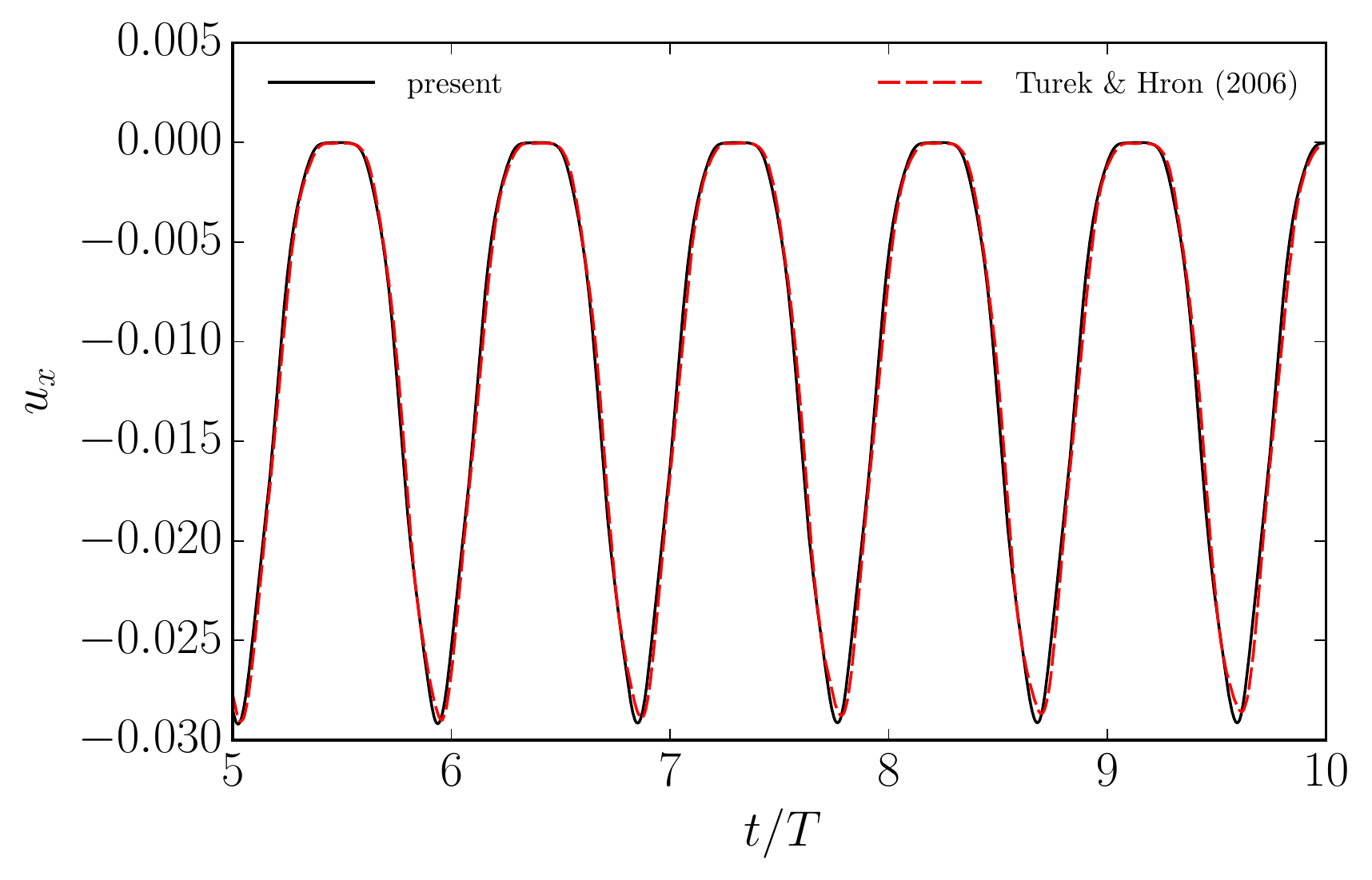}
                \includegraphics[width=0.45\textwidth]{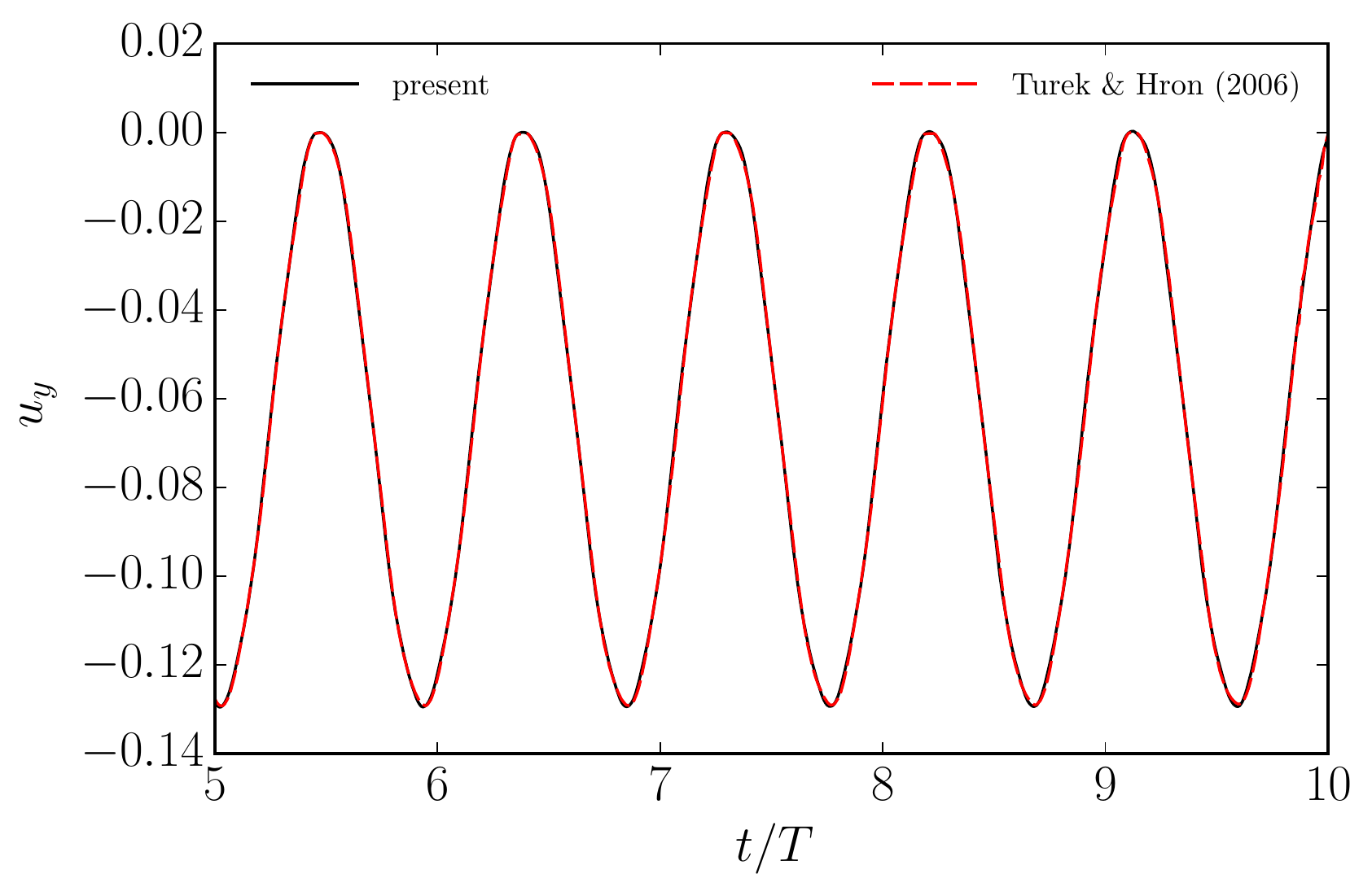}
        \caption{CSM3: $x$- and $y$-displacement of the beam tip.}
	\label{fig:csm3}
\end{figure*}

\begin{table}
\centering
\begin{tabular}{lccc}
\hline
Contribution                          & $u_x$                       &$u_y$                      &$f$\\
\hline
\cite{Turek2006}                      & $-0.01431  \pm 0.01431$     & $-0.06361 \pm 0.06516$    &$1.0995$  \\ 
present                               & $-0.01460  \pm 0.01460$     & $-0.06463 \pm 0.06492$    &$1.10$ \\
\hline
\end{tabular}
\caption{Results for CSM3.}
\label{tab:sd7003_lsb_props}
\end{table}

\begin{figure*}
	\centering					
	\includegraphics[width=0.7\textwidth]{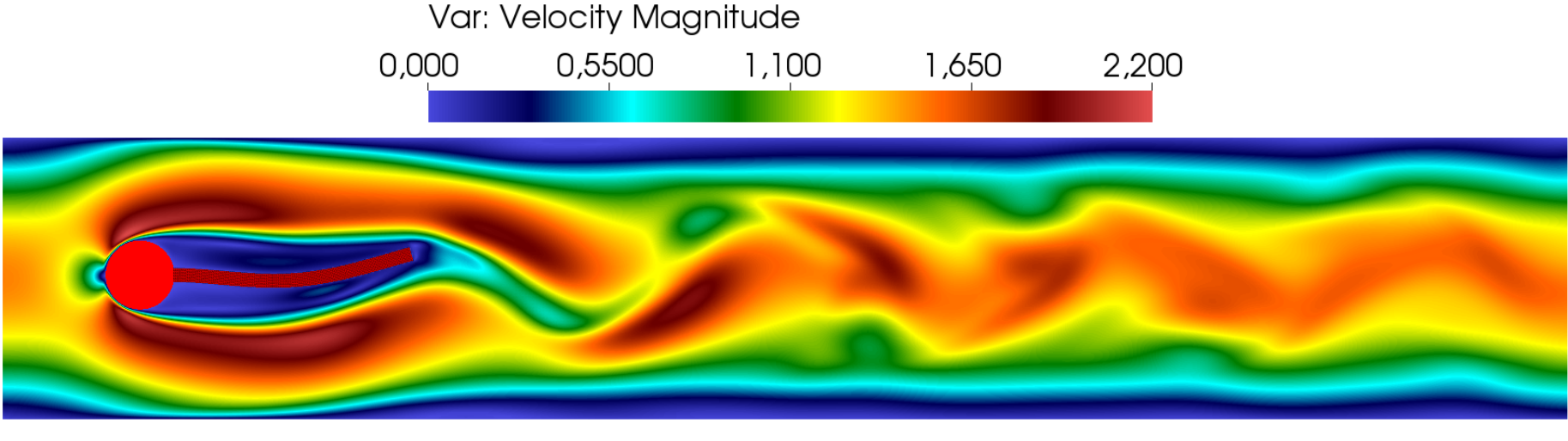}
	\caption{FSI3: Snapshot of velocity magnitude.}
	\label{fig:Turel_velMag}
\end{figure*}

\begin{table}
\centering
\begin{tabular}{lcccc}
\hline
Contribution                                            & $u_x$                      &$u_y$                       &$f_x$    &$f_y$ \\
\hline
LB-FEM\citep{Kollmannsberger2009}           & $-0.00288  \pm 0.00271$    & $0.00148 \pm 0.0351$       &$11$     &$5.5$\\ 
ALE-FEM \citep{Turek2006}                   & $-0.00269  \pm 0.00253$    & $0.00148 \pm 0.03438$      &$10.9$   &$5.3$\\ 
present                                     & $-0.00268  \pm 0.00257$    & $0.00145 \pm 0.03380$      &$11$    &$5.5$\\
\hline
\end{tabular}
\caption{FSI3: Mean and amplitude of the flag tip deflection.}
\label{tab:turek_fsi}
\end{table}

\begin{figure*}
        \centering
                \includegraphics[width=0.45\textwidth]{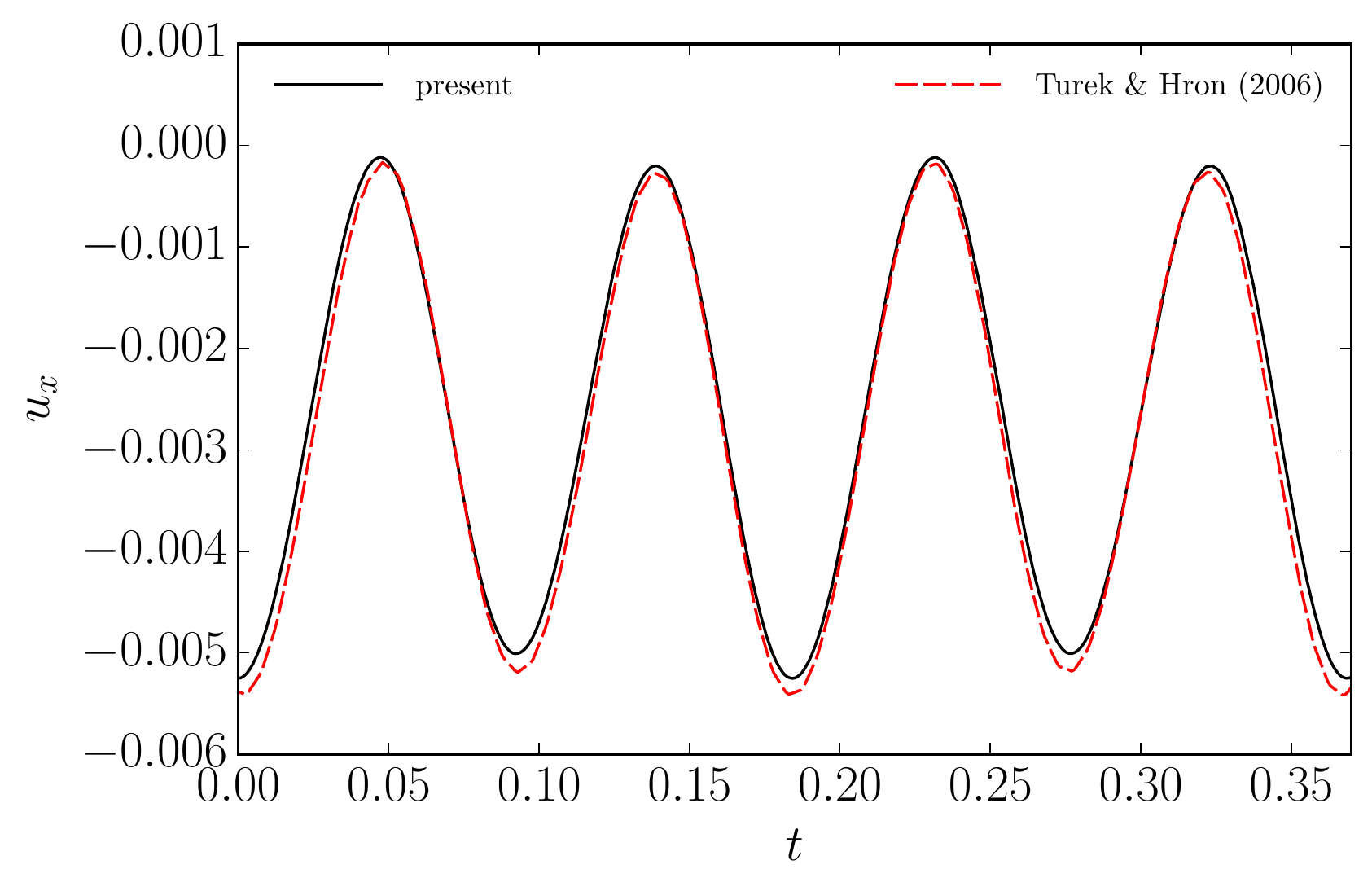}
                \includegraphics[width=0.45\textwidth]{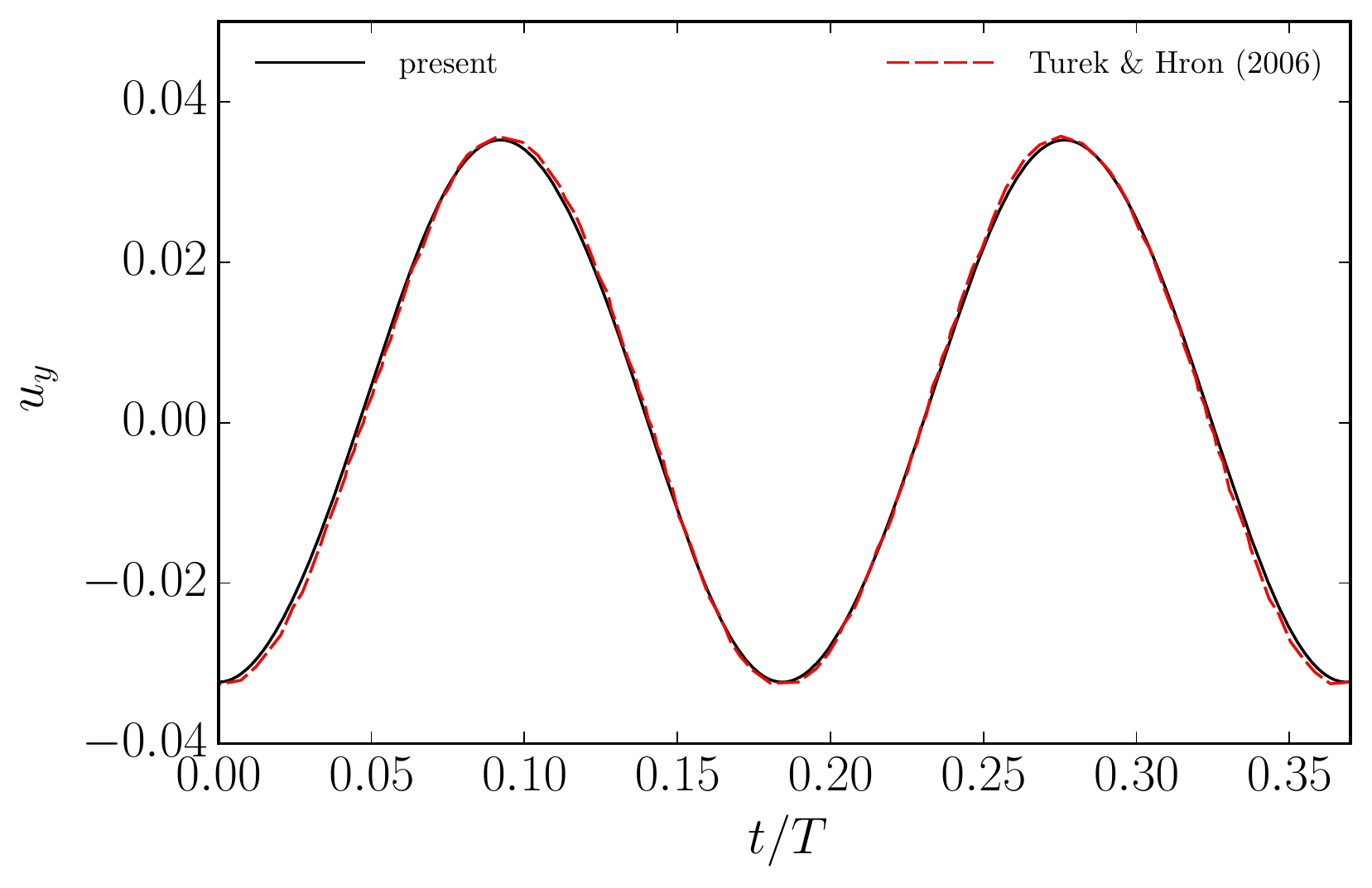}
    \caption{FSI3: Evolution of the flag tip deflection in $x$- and $y$-direction.}
	\label{fig:evol_displ_fsi3}
\end{figure*}

For the validation of FSI schemes a comprehensive test suite was proposed in
\cite{Turek2006}, which consists of a rigid circular cylinder with a flexible
flag attached to its downstream side.  The structure is placed asymmetrically in
a laminar channel flow and therefore induces an oscillatory motion of the
elastic beam as the flow evolves.  The setup is schematically shown in
Figure~\ref{fig:csm3_schematic}.  While the boundary conditions in pitchwise
directions are no-slip boundaries, the inflow at the left boundary has a
prescribed parabolic velocity profile according to \begin{equation} u(0,y)= 1.5
\bar{U} \frac{y(H-y)}{(H/2)^2}, \end{equation} where the mean inflow velocity is
$\bar{U}$ and the channel height $H$.  As initial condition for the unsteady
simulation we use a smooth ramping function for the inflow.  The cylinder with
diameter $D$ is placed asymmetrically at $(2D,2.1D)$, while the beam has length
$L=3.5D$ and thickness $h=0.2D$.  Note that while in \cite{Turek2006} all
computations were carried out in two dimension, we perform a quasi
two-dimensional simulation by using only a few points in spanwise direction and
apply periodic and plane strain boundary conditions for the fluid and the solid,
respectively.  The constitutive law for the solid part is assumed to follow the
hyperelastic Saint Venant-Kirchoff model.

Before attempting to solve the fully coupled FSI system, we first validate the
structural solver separately using a time-dependent large deformation test case.
Thus, we do not consider the surrounding fluid of the setup in
Figure~\ref{fig:csm3_schematic}, but only account for a gravitational force
$\textbf{g} =(0, 2 \cdot 10^3 \rm{m/s^2} )$, which is acting on the beam with
density $\rho_s= 10^3\rm{kg/m^3}$. The Poisson ratio and the shear modulus are
taken as $\nu_s=0.4$ and $\mu_s=0.5 \cdot 10^6 \rm{kg/ms^2}$, respectively.
This corresponds to CSM3 in \cite{Turek2006}, where the authors report the
evolution of the beam tip displacement in $x$ and $y$ direction. For this setup
the beam was discretized by $280$ elements and evolved using a time step of
$\Delta t= 0.001$.  The comparison to \cite{Turek2006} is shown in Figure
\ref{fig:csm3} and it is obvious that apart from minor artificial damping in the
simulations of \cite{Turek2006} both results agree well. This validates our
implementation of the structural model.

For brevity, we avoid presenting the pure CFD validation as done
in \cite{Turek2006}. The fluid solver however was thoroughly validated as
witnessed by many of our preceding contributions (see, e.g.,
\cite{dorschner2016grid}).  Thus, having validated the structural solver, we
proceed with benchmarks of the fully coupled FSI scheme.  To that end, we
consider the FSI3 benchmark of \cite{Turek2006} for which the density ratio is
$\rho_s/\rho_f=1$ and the Reynolds number $Re= \bar{U}D/\nu=200$.  The
aeroelastic coefficient was taken as $\rm{Ae}=E_s/(\rho_f \bar{U}^2)=1.4 \cdot
10^3$, where $E_s$ indicates Young's modulus for the structure and the Poisson
ratio was set to $\nu_s=0.4$.  In the fluid domain, we use two levels of
refinement as shown in Figure \ref{fig:TurekMesh}, which effectively resolves
the cylinder diameter by $D_{lb}=40$ lattice points.  The elastic beam was
discretized using $(140,10,1)$ elements and evolved with a time step of $\Delta
t=8.75\cdot 10^{-5}$.

On one hand, the elastic beam is periodically excited by the vortex street in
the wake of the cylinder, which yields strongly nonlinear deflections of the
flag. On the other hand, the momentum transferred from the solid excites the fluid.
A slice through the computational domain, plotting a snapshot velocity magnitude
is shown in Figure \ref{fig:Turel_velMag}.  More quantitatively, we computed the
mean and amplitude of the deflection at the free end of the flag along with the
corresponding oscillation frequencies.  The comparison with literature values is
excellent and listed in Table \ref{tab:turek_fsi}.  Finally, the deflection
evolution is reported in Figure \ref{fig:evol_displ_fsi3}, which agrees well
with the reference data in \cite{Turek2006}.

\subsubsection{Flow past a flapping flag}

\begin{figure}
        \centering
                \includegraphics[width=0.3\textwidth]{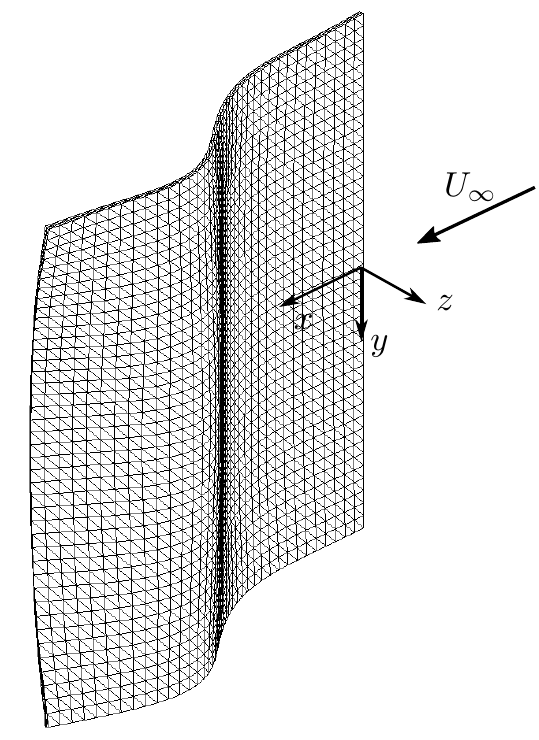}
        \caption{Schematic of the flow past a flapping flag}
	\label{fig:schematic_flag}
\end{figure}

\begin{figure*}
        \centering
            \includegraphics[width=0.45\textwidth]{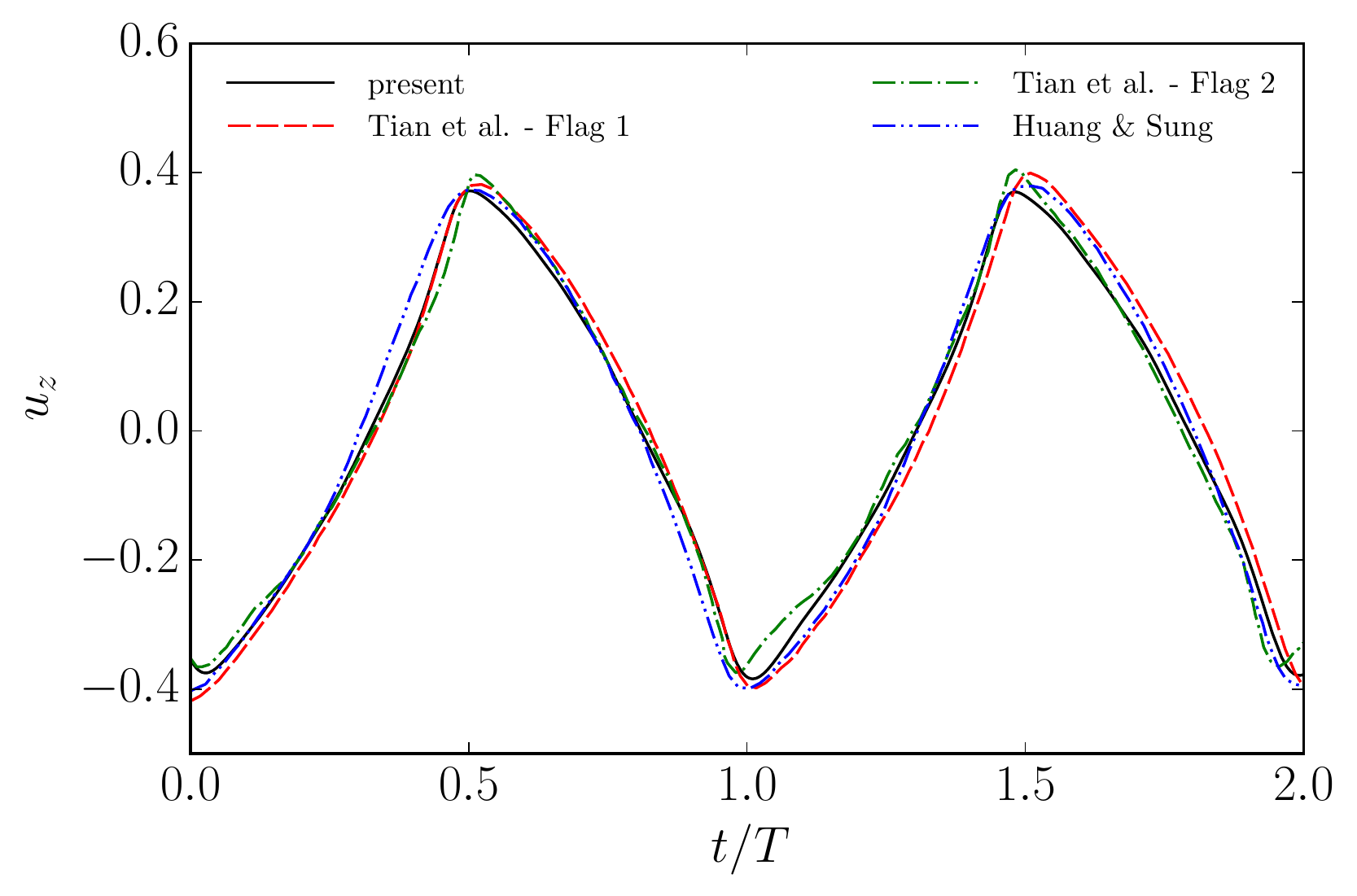}
            \includegraphics[width=0.45\textwidth]{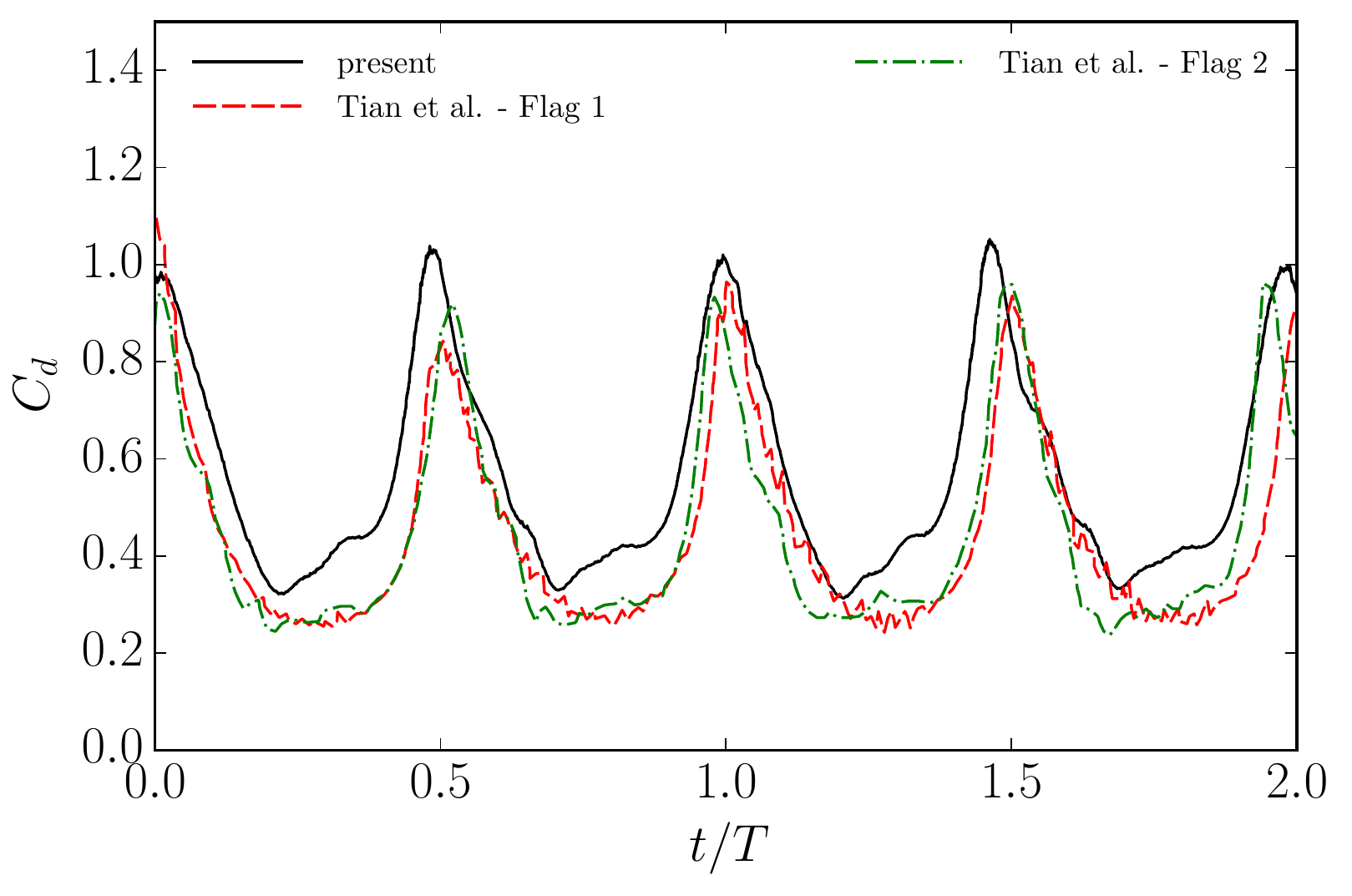}
            \caption{Flow past a flapping Flag. Left: Displacement of the point B
                located at $B=(L,0,0)$ in the undeformed configuration. Right: Evolution of the drag coefficient.}
	\label{fig:displ_flag}
\end{figure*}

\begin{figure}
	\centering					
	\includegraphics[width=0.35\textwidth]{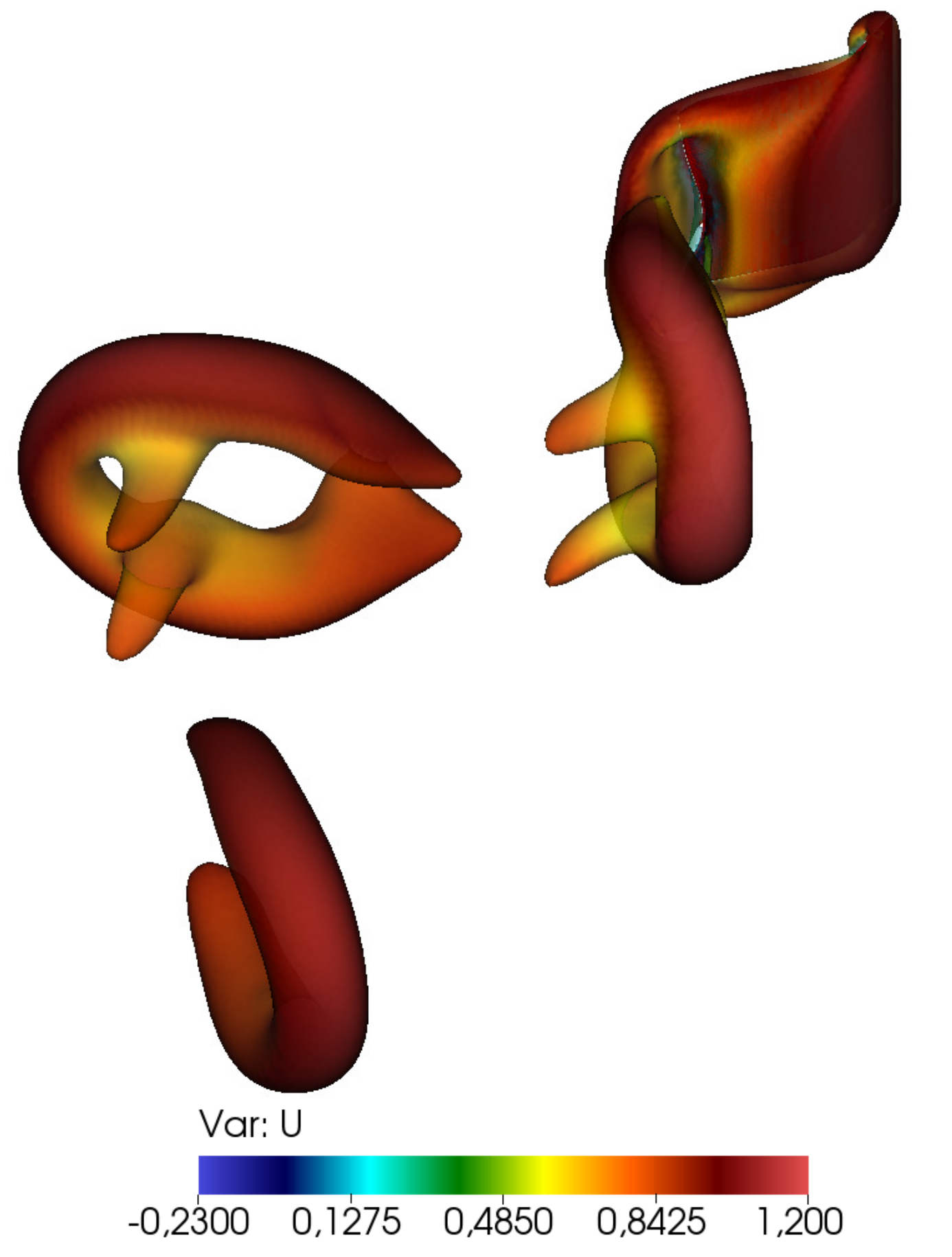}
	\caption{Wake of a flapping flag visualized by isosurfaces of Q-criterion
        and colored by streamwise velocity.}
	\label{fig:qIsosurface_flag}
\end{figure}

Having validated the proposed scheme in the quasi two-dimensional setting, we
next consider a fully three-dimensional flow.  To that end, we investigate the
non-linear dynamics of a flag in a uniform fluid flow. Despite being a classical
model problem for FSI, the complex motion of the flag challenges numerical
methods and thus only a few cases have been reported in literature
\cite{Tian2014,Huang2010,Kim2007}.  Here, we use the case as provided in
\cite{Tian2014,Huang2010} for validation.  While in \cite{Huang2010} a
diffuse-interface immersed boundary method was employed, \cite{Tian2014} used an
immersed boundary method coupled with a nonlinear FEM solver.  As shown in
Figure \ref{fig:schematic_flag}, the leading edge of a square flag of length $L$
and thickness $h=0.01L$ is placed at the origin of the domain. In addition, zero
displacement and velocity boundary conditions are imposed at the leading edge.
The rectangular fluid domain spans from $[-2L \times-1L \times -4L]$ to $[8L
\times 1L \times 4L]$ in streamwise, spanwise and transverse direction,
respectively, where periodic boundary conditions are applied in spanwise
direction and free-stream boundaries are imposed in transverse direction.  Using
two levels of refinement, the flag was resolved by $L=100$ lattice units in the
finest level.  The flexible flag is discretized with a uniform mesh of $[50
\times 50 \times 2]$ elements.  The Poisson ratio is set to $\nu_s=0.4$ and the
bending rigidity is $Eh^3/(12(1-\nu_s^2)\rho_f U_\infty^2 L^3)=10^{-4}$.  The
density ratio is taken as $\rho_s/\rho_f =L/h$ and the Reynolds number is
$Re=U_\infty L/\nu=200$.  Initially, the flag coincides with the $xy$-plane and
a small perturbation is used to trigger the periodical flapping behavior.
During the evolution, we record the displacement of the Point B, which is
located at $B=(L,0,0)$ in the undeformed configuration and compare it to the
reference data of \cite{Tian2014, Huang2010} in Figure \ref{fig:displ_flag}.
Note that in \cite{Tian2014} two flag models were considered, namely a plate
model with infinitesimal thickness (Flag 1) and three dimensional model with
thickness $h=0.01L$ (Flag 2). Both cases demonstrate negligible discrepancies
due to the low Reynolds number in this case.  After the initial transient, the
flow quickly converges to a periodic flapping as seen in Figure
\ref{fig:displ_flag}. The comparison of the present simulations to the
references shows good agreement.  Besides the displacement, we computed the
evolution of the drag coefficient $C_d=F_x/(1/2  \rho_f U_\infty^2 L^2 )$ and
compare it values reported in \cite{Tian2014} in Figure \ref{fig:displ_flag}.
Significant noise can be observed for the simulations by \cite{Tian2014}, which,
according to the authors, originates from the noisy prediction of the thin
plate. In contrast, the results from the present method appear smooth and do not
exhibit oscillations. Overall, both results agree qualitatively but do exhibit
discrepancies, likely due to the noise.
Unfortunately, no data regarding the drag evolution was reported in
\cite{Huang2010} and thus eludes a comparison.
Finally, in Figure \ref{fig:qIsosurface_flag}, the vortical structures in the
wake of the flag are visualized by isosurfaces of the $Q$-Criterion, which are
colored by streamwise velocity.  The vortices shed from the trailing edge
connect with the vortices shed from the side edges to form hairpin-type vortices
along with two separate co-rotating vortices. Notably, this wake structure bears
significant resemblance to self-propelled anguilliform swimmers
\cite{dorschner2017entropic}. Analogous vortex structures have been observed in
the references \cite{Tian2014, Huang2010}.

\subsubsection{Beam in crossflow}

So far we have successfully validated the proposed scheme for quasi-two
dimensional and three-dimensional flows. As a final validation, we include a
simulation involving turbulence. To that end, we consider a flexible beam in a
cross flow.  This set-up has been studied both experimentally and numerically in
\cite{Tian2014} and \cite{Luhar2011}, respectively and aims to model the
deformation of aquatic plants caused by the flow.  The beam is vertically
mounted in a uniform flow and has the length $L$, the thickness $h$ and the
width $b$.  As in the references, the Reynolds number is set to $Re=U_\infty
L/\nu=8000$ and the geometrical properties of the beam are given by $L/b=5$
$h/b=0.2$.  The solid material has the non-dimensional Young's modulus
$\tilde{E}_s=E_s/\rho U_\infty^2$ and the Poisson's ratio $\nu_s=0.4$. The
density ratio is set to $\rho_s/\rho_f=0.678$ and a buoyancy force
$f_b=(\rho_f-\rho_s)g h / (\rho_f U_\infty^2) =0.2465$ is applied.  The
rectangular domain ranges from $[-5b, -8b, -8.5b]$ to $[16b, 8b, 8.5b]$ in which
the centroid of the beam is placed at the origin of the undeformed
configuration.  Using one level of refinement, the fluid domain discretizes the
beam width with $b=40$ lattice points and the solid mesh employs $[2 \times
20\times140]$ elements to represent to beam.

\begin{figure}
	\centering					
	\includegraphics[width=0.45\textwidth]{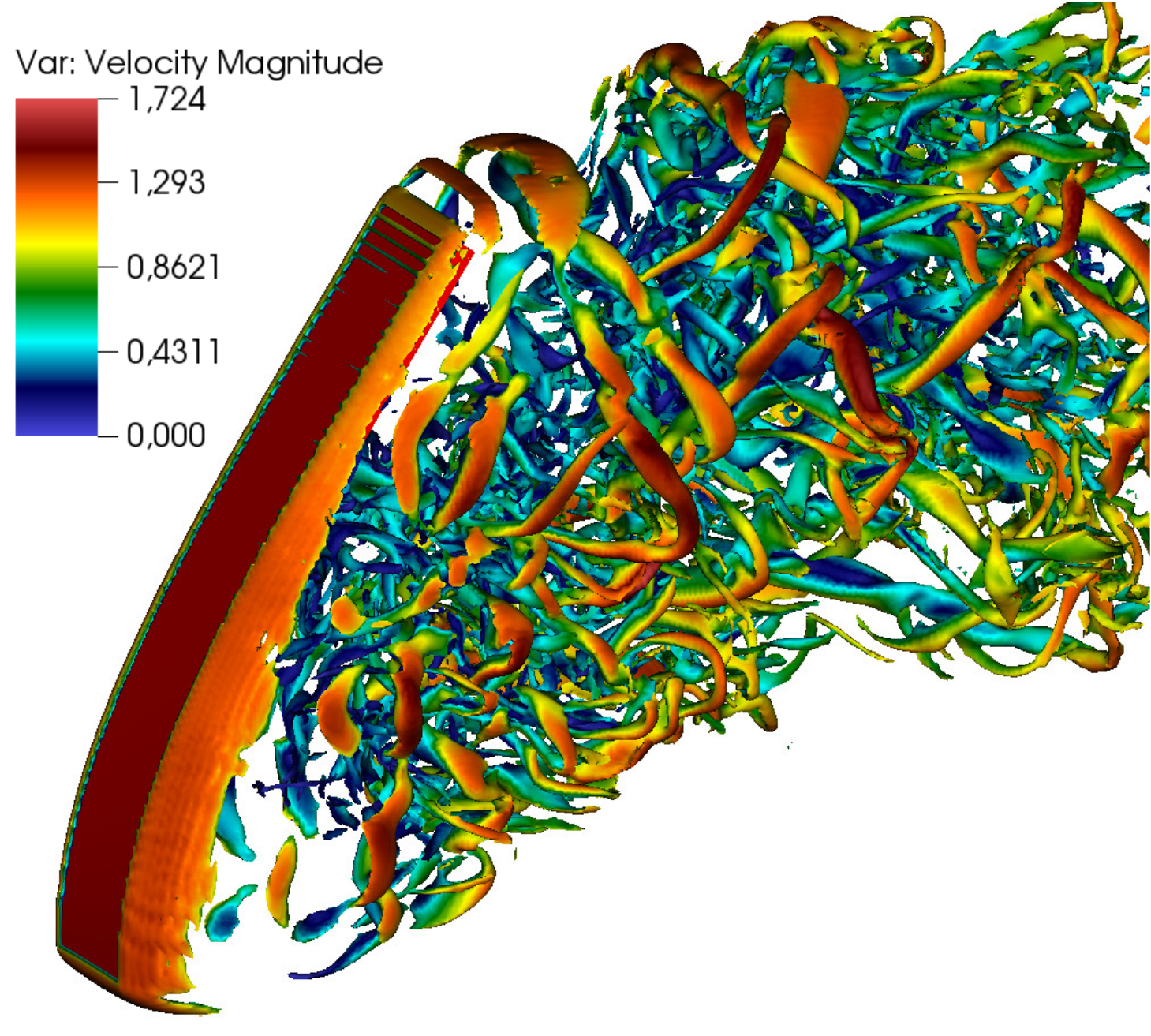}
    \caption{Flexible plate in a cross flow: Isosurfaces of Q-criterion, colored
    by velocity magnitude.}
	\label{fig:snapshot_CrossFlow}
\end{figure}

Using these flow and structural conditions and parameters, the plate converges
to a steady deformation. A snapshot of the deformed state is presented in Figure
\ref{fig:snapshot_CrossFlow}, where the wake behind the deformed beam is
visualized by isosurfaces of the Q-Criterion and colored by velocity magnitude.
Qualitatively this is in line with the reference.  For a more thorough
comparison, we computed the drag coefficient $C_d=F_x/(1/2 \rho_f U_\infty ^2 b
L)$ along with the deflection of the beam's free end in the deformed state.
Along with the reference values, the results of the present simulation are
listed in Table \ref{tab:fsi_CrossFlow}.  It is apparent that the results are in
good agreement with the reference data.  While some discrepancies may be
observed to the numerical study of \cite{Tian2014} the present simulation
matches the experimental study well \cite{Luhar2011}.

\begin{table}
\centering
\vspace{1cm}
\begin{tabular}{lccc}
\hline
Contribution                    & $C_d$                     &$D_x/b$          &$D_z/b$       \\
\hline
Tian et al. (2014), IMB-FEM     & $1.03$                    & $2.12$          &$0.54$        \\ 
Luhar \& Nepf (2011), exp.      & $1.15$                    & $2.14$          &$0.59$        \\ 
present                         & $1.13$                    & $2.14$          &$0.55$       \\
\hline
\end{tabular}
\caption{Flexible plate in a cross flow.  Comparison of drag coefficient $C_d$
and plate deflection $D_x/b$ and $D_y/b$ in stream- and pitchwise direction,
respectively.}
\label{tab:fsi_CrossFlow}
\end{table}

\subsection{Extensions to fluid-structure interaction in multiphase flow}
\label{sec:fsi_mpf}

Two-phase flows are of fundamental interest in science and engineering
applications~\cite{Brennen2005}, which exhibits various complex phenomena at
multiple temporal and spatial event scales \cite{yarin2006drop,marengo2011drop}.
These include droplet breakup, droplet reconnection as well as droplet impact on
a surface, where effects such as splash
\cite{mandre2012mechanism,riboux2014experiments},
skating\cite{kolinski2012skating,de2015wettability}, rebound
\cite{richard2002surface,bird2013reducing,liu2014pancake} or the trampoline
effect \cite{schutzius2015spontaneous} have been observed.

In recent years, much attention has been devoted to droplet impact on so-called
superhydrophobic surfaces. Super-hydrophobic surfaces exhibit strong repellence
of liquid droplets, which can be exploited in for anti-icing, self-cleaning,
drag reduction and many other applications
\cite{bhushan2011natural,blossey2003self,Kreder2016}.  The most known example of
a natural superhydrophobic surface is the surface of the lotus leaf, i.e.
Nelumbo nucifera. Numerous studies suggested that the combination of surface
chemistry and roughness on multiple scales on the surface is responsible for its
repellence.  Thus, modern synthetic designs of superhydrophobic surfaces,
combine the effects of micro-texturing and chemistry to enhance the hydrophobic
effect.  To that end, many studies have investigated the underlying physics of
droplet impact on superhydrophobic surfaces using different designs and
conditions with the ultimate goal to reduce the contract time
\cite{schutzius2014physics,Bird2002}.  Note however that most studies have
focused on rigid surfaces and neglected the flexibility of the substrate, which
is inherent to most naturally occurring repellent surfaces such as leaves,
textiles or butterfly wings.  Notable is the recent study of
\cite{Vasileiou2016}, where the effect of elasticity on hydrophobicity was
investigated experimentally.

In this section, we aim to go beyond classical benchmark cases and explore the
capabilities of the KBC-FSI solver in the context of multiphase flows by considering
droplet impact on flexible superhydrophobic surfaces, similar to
\cite{Vasileiou2016}.  From the numerical point of view, simulations of such
a kind are challenging.  However, the LBM offers an attractive alternative to
conventional schemes, due to the ease of  implementing inter-molecular forces
and complex boundaries without sacrificing  efficiency \cite{Mazloomi2015a}.
While various LB models for multiphase flow exist, restrictions on density
ratio, kinematic viscosity and interface thickness remained for long.  Among
others, a viable alternative was proposed in \cite{Mazloomi2015a}, where
combining the notion of a discrete entropy function, the free-energy based
formulation and an appropriately regularized equation of state significantly
increased the range of applicability of LB models for multiphase flow.  This
approach has been thoroughly validated by simulations of droplet impact on flat
and micro-textured superhydrophobic surfaces for a variety of different bouncing
regimes \cite{MazloomiM2015,Moqaddam2016b}.  Here, we build on these results and
extend it to the KBC model, the Grad boundary condition and the coupling to the
structural solver. The equation of state and the forcing approach is kept same.

On the fluid side, following \cite{Mazloomi2015a}, the phase separation and
wetting properties are implemented through a body force 
\begin{equation}\label{eq:Force}
\bm F = \bm F_f + \bm F_s.
\end{equation}
The mean field force 
\begin{equation}\label{eq:force1}
F_{f,\alpha}= \partial_\beta \left( \rho c_s^2 \delta_{\alpha\beta} - P_{\alpha \beta}^{\rm K} \right),
\end{equation}
accounts for the phase separation by 
implementing the Korteweg stress tensor 
\begin{equation}\label{eq:Korteweg}
P_{\alpha\beta}^{\rm K} = \left(p -\kappa \rho \partial_\gamma \partial_\gamma
        \rho - \frac{\kappa}{2} (\partial_\gamma \rho)(\partial_\gamma \rho)
        \right) \delta_{\alpha \beta} + \kappa (\partial_\alpha
            \rho)(\partial_\beta \rho),
\end{equation}
where the pressure $p$ is prescribed through a non-ideal equation of state and
$\kappa$ controls the surface tension. 
This yields
\begin{equation}\label{eq:force2}
F_{f,\alpha} = 2\varphi \partial_\alpha \varphi - \kappa \rho \partial_\alpha \left( \partial_\beta \partial_\beta \rho \right),
\end{equation}
with 
\begin{equation}
\varphi = \sqrt{\rho c_s^2 - p}.
\end{equation}

The equation of state is a polynomial regularization of 
Peng-Robinson form~\cite{Yuan2006} as introduced in~\cite{Mazloomi2015a} and
reads 
\begin{equation}
\begin{split} p = 
&5.3\cdot 10^{-2}\rho \\
&-3.818183621928911\cdot 10^{-2} \rho^2 \\
&+4.139745482116095\cdot 10^{-3} \rho^3 \\
&+ 3.748484095210317\cdot 10^{-4} \rho^4 \\
&- 1.4552652965531227 \cdot 10^{-4} \rho^5 \\
&+ 1.2746947442749278\cdot 10^{-5} \rho^6,
\end{split}
\label{eq:eos}
\end{equation}
which yields an effective density ratio of $\rho_v/\rho_l \approx 100$ with
liquid and vapour densities $\rho_l\approx 7.55$ and $\rho_v\approx 0.073$,
respectively.

Different wetting states can be modeled by means of
the force $\bm F_s$, which reads
\begin{equation}
F_{s,\alpha}(x_\alpha,t)= \kappa_w \rho(x_\alpha,t) \sum_i^N w_i s(x_\alpha+
        c_{i,\alpha}\delta t) c_{i,\alpha},
\end{equation}
where $\kappa_w$ allows to us to choose the equilibrium contact angle in
accordance with the Young-Laplace equation.  The term $s(x + c_{i, \alpha}
\delta t)$ is an indicator function that is equal to one for the solid domain
nodes and is equal to zero otherwise; $w_i$ are appropriately chosen weights
\cite{Mazloomi2015a}.  To model superhydrophobic surfaces, the equilibrium
contact angle was set to $\theta=165^\circ$, which corresponds to
$\kappa_w=-0.145$.

The total body force $\bm F$ is imposed through the exact difference method
\cite{Kupershtokh2004} with the velocity increment
\begin{equation}
\delta u_\alpha  = \frac{F_\alpha}{\rho \delta t}.
\end{equation}
Hence, the LB equation can be written as
\begin{equation}
f_i(\bm{x}+\bm{v}_i,t+1)=f_i'\equiv(1-\beta)f_i(\bm{x},t)+\beta \fmirr_i(\bm{x},t) + F_i(\bm{x},t),
\end{equation}
with
\begin{equation}\label{eq:force_mp}
F_i = \feq_i(\rho,\bm u+\delta \bm u) - \feq_i(\rho,\bm u).
\end{equation}
Unlike, the entropic LBM of \cite{Mazloomi2015a}, we here use the KBC
realization of LBM, where we incorporate the force term into the KBC model
through the shifted entropic scalar product
\begin{equation}
\edotshift{X}{Y} = \sum_i \frac{X_i Y_i}{\feq(\rho,\bm u+\delta \bm u)},
\end{equation}
which is used to compute the stabilizer $\gamma$ from Eq.~\eqref{eq:gamma_min_approx}. 

Also in the multiphase model, we use Grad's boundary condition.  For the
fluid-structure coupling, we employ the same methodology as outlined above but
include the pressure $p$ as prescribed by the equation of state in
Eq.~\eqref{eq:eos}.  Note however that the diffuse nature of the liquid-vapour
interface necessitates a pressure regularization. This arises from the fact that
the numerical integration of the pressure over the solid surface is prone to
numerical errors, due to sharp pressure gradients and large negative values in
the interface region, which are sampled only relatively coarsely on the FEM
mesh.  This leads to an artificial negative pressure, which is compensated in
our simulations by a regularization procedure, where we use a simple linear
interpolation between the liquid and vapor density to evaluate the pressure.

Motivated by the experimental study of \cite{Vasileiou2016}, we
investigate the effect of elasticity on the droplet impact on a
superhydrophobic, elastic beam for a wide range of Weber numbers.

In all simulations, the droplet is resolved by $D=80$ lattice points, the
surface tension is set to $\sigma=0.295$ $(\kappa=0.295)$ and the computational
domain of the fluid is given by $[320 \times 250 \times 320]$.  The beam has
dimensions $[300 \times 200 \times 5]$ and Lam\'e coefficients are set to
$\lambda_s=1500$ and $\mu_s=1000$. While one end of the beam is clamped, the
other end is only simply supported and the droplet impacts the center of the
beam.

\begin{figure}
	\centering					
	\includegraphics[width=0.45\textwidth]{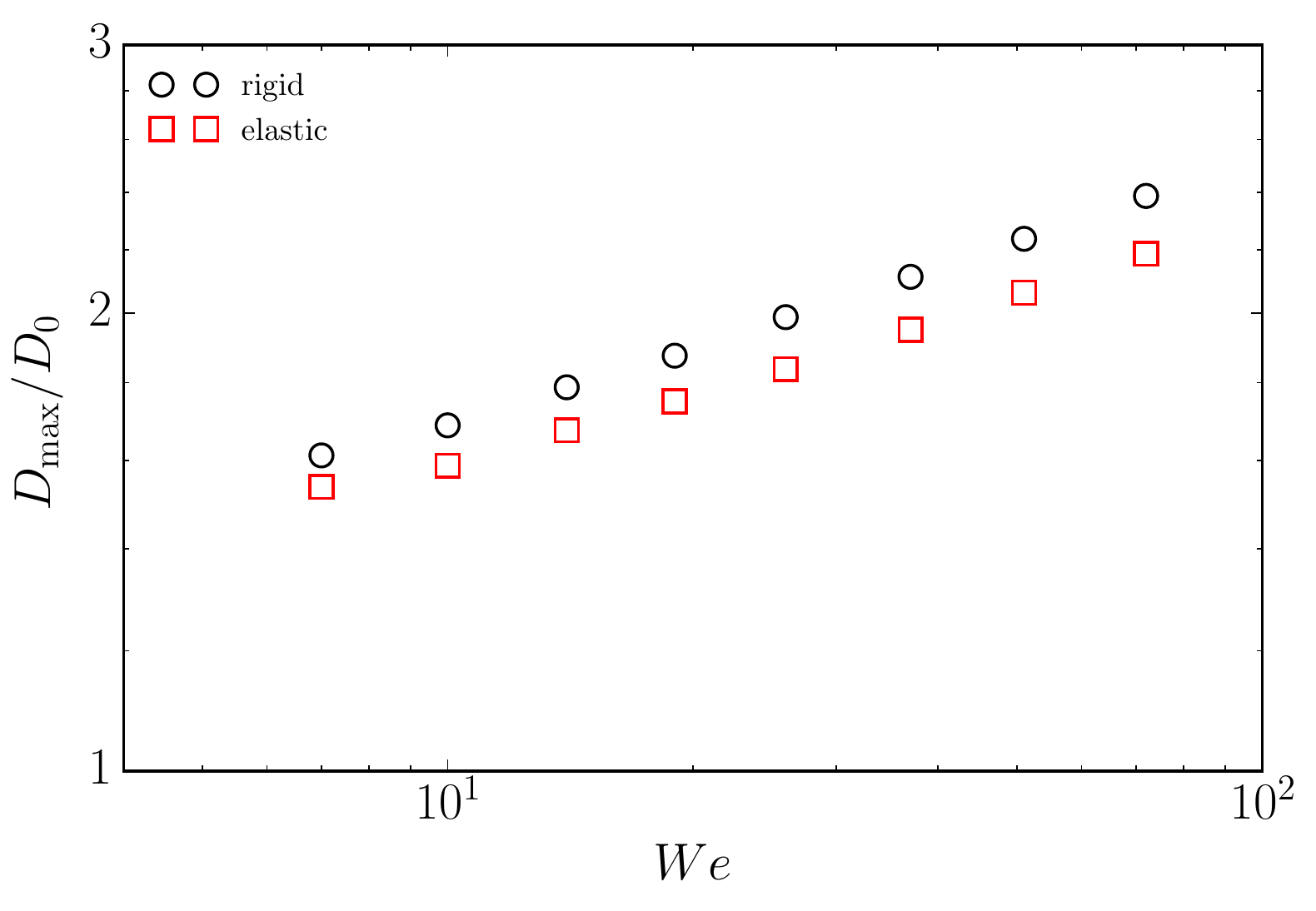}
    \caption{Maximum droplet spreading diameter on a rigid and elastic
    superhydrophobic surface, respectively.}
	\label{fig:spreading_diameter_rigid_vs_elastic}
\end{figure}

We simulated Weber numbers in the range of $We \in [7, 72]$ for both rigid and
flexible beams and recorded the maximum spreading diameter $D_{max}/D_0$ as
shown in Figure \ref{fig:spreading_diameter_rigid_vs_elastic}.

\begin{figure*}
	\centering					
	\includegraphics[width=1\textwidth]{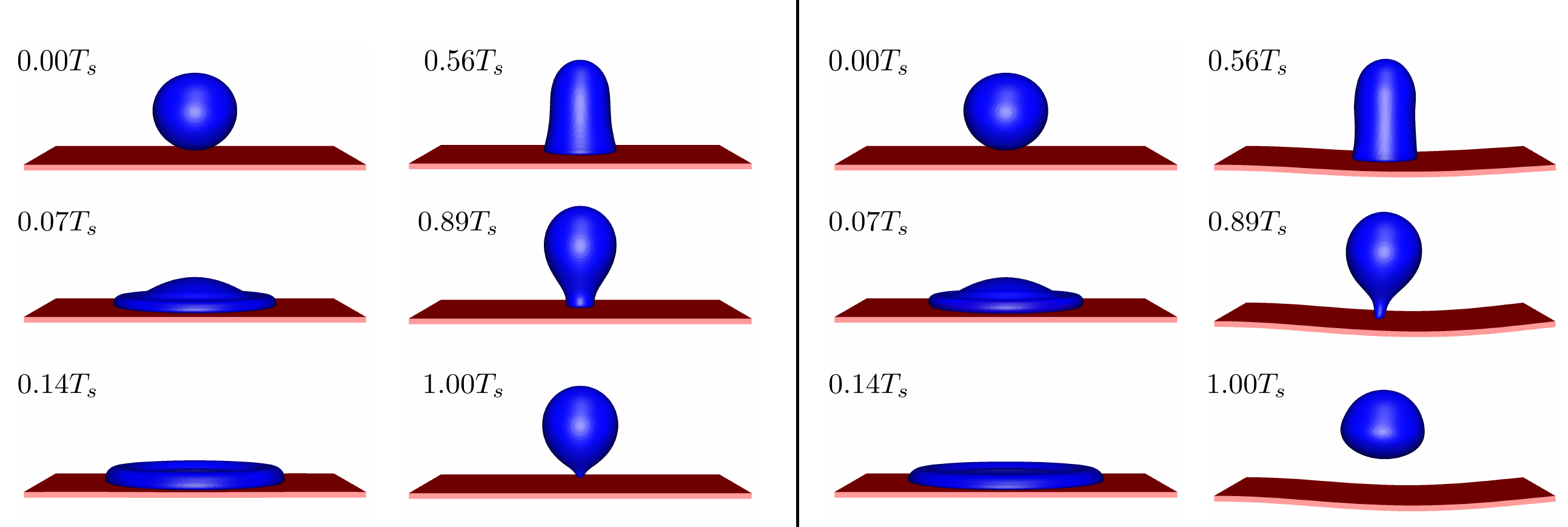}
	\caption{Droplet spreading on a rigid (left) and elastic (right)
        superhydrophobic surface. Timings are normalized by the contact time
            $T_s$ of the rigid surface.}
	\label{fig:droplet_snapshots}
\end{figure*}
For the entire range of Weber numbers, it is apparent that the maximum spreading
diameter decreases when elasticity of the beam  is taken into consideration.
Analogously, the experimental study conducted in \cite{Vasileiou2016} also
observed a reduction of the apparent spreading diameter.  {While a
quantitative comparison is out of reach for the current preliminary simulations
due to the large dimensions of the beam used in the experiment, the 
proposed scheme does capture the effect of elasticity qualitatively.
A natural explanation for the cause of the reduction of the spreading diameter is
that the momentum of the droplet is transferred to the beam, which decreases the
effective Weber number perceived by the droplet and thus reduces the maximum
spreading of the droplet.  It is only long after the droplet has reached its
maximum spread that the momentum is transferred back (no damping is applied) to
the liquid.  A similar explanation was proposed in \cite{Vasileiou2016}.  A
sequence of snapshots of the droplet impact on both the rigid and the elastic
beam is shown in Figure \ref{fig:droplet_snapshots}.  It is clear that initially
both the rigid and the elastic beam behave similar, but the elastic case
exhibits faster rebound and take off. Note that the density ratio between solid
and fluid is roughly $\rho_s/\rho_f \approx 100$, which explains the delayed
response of the fluid.  Further, the observed asymmetry in the elastic case is
due to the asymmetric boundary conditions of the beam.

These results are promising and underline the robustness and viability
of multi-physics simulations based on the KBC-FSI solver. 
A detailed investigation of FSI for multiphase flows will be published in a
subsequent paper.

\section{Concluding remarks}
In this paper we have presented a partitioned fluid-structure interaction
approach. On one hand, the fluid flow is computed by the entropic
multi-relaxation time lattice Boltzmann model in combination with Grad boundary
conditions and multi-domain grid refinement.  On the other hand, the elastic
solid was modeled by the hyperelastic Saint Venant-Kirchoff model, which
accounts for large, geometrically non-linear deformations and was solved by a
corresponding FEM formulation.

The proposed scheme was validated for various challenging set-ups for quasi-two
dimensional and fully three dimensional simulations of laminar and turbulent
flows.  Finally, extensions to multi-physics simulations were explored. 
An extension of the KBC model to multiphase flows and its coupling to the solid
solver was presented. Promising results, in qualitative agreement with recent
experiments, were shown for the simulation of droplet impact on elastic
superhydrophobic surface, which demonstrate the viability of proposed scheme.

\begin{acknowledgments}
This work was supported by the European Research Council (ERC) Advanced Grant
No. 291094-ELBM, the ETH-32-14-2 grant and the SNF grant 200021\_172640.  The
computational resources at the Swiss National Super Computing Center CSCS were
provided under the grants s630 and s800.
\end{acknowledgments}

\bibliography{library}

\end{document}